**Phase Angle Effects on 3-μm Absorption Band on Ceres: Implications for Dawn Mission**

D. Takir
Planetary Science Institute, 1700 E Fort Lowell Road, Suite 106, Tucson, AZ 85719, USA
Department of Physics and Astronomy, Ithaca College, Ithaca, NY 14850, USA
dtakir@psi.edu.

V. Reddy
Planetary Science Institute, 1700 E Fort Lowell Road, Suite 106, Tucson, AZ 85719, USA

J. A. Sanchez
Planetary Science Institute, 1700 E Fort Lowell Road, Suite 106, Tucson, AZ 85719, USA,

L. Le Corre

Planetary Science Institute, 1700 E Fort Lowell Road, Suite 106, Tucson, AZ 85719, USA

P. S. Hardersen
Department of Space Studies, University of North Dakota, Grand Forks, ND 58202, USA.

A. Nathues
Max-Planck-Institute for Solar System Research, Justus-von-Liebig-Weg 3, 37077 Göttingen, Germany

# ABSTRACT

Phase angle-induced spectral effects are important to characterize since they affect spectral band parameters such as band depth and band center, and therefore skew mineralogical interpretations of planetary bodies via reflectance spectroscopy. Dwarf planet (1) Ceres is the next target of NASA's Dawn mission, which is expected to arrive in March 2015. The visible and near-infrared mapping spectrometer (VIR) onboard Dawn has the spatial and spectral range to characterize the surface between 0.25-5.0 µm. Ceres has an absorption feature at 3.0 µm due to hydroxyl- and/or water-bearing minerals (e.g. Lebofsky et al. 1981, Rivkin et al. 2003). We analyzed phase angle-induced spectral effects on the 3-µm absorption band on Ceres using spectra measured with the long-wavelength cross-dispersed (LXD: 1.9-4.2-µm) mode of the SpeX spectrograph/imager at the NASA Infrared Telescope Facility (IRTF). Ceres LXD spectra were measured at different phase angles ranging from $0.7^o$ to $22^o$. We found that the band center slightly increases from 3.06 µm at lower phase angles ($0.7^o$ and $6^o$) to 3.07 µm at higher phase angles ($11^o$ and $22^o$), the band depth decreases by ~20% from lower phase angles to higher phase angles, and the band area decreases by ~25% from lower phase angles to higher phase angles. Our results will have implications for constraining the abundance of OH on the surface of Ceres from VIR spectral data, which will be acquired by Dawn starting spring 2015.

# 1. Introduction

The major goal of the Dawn mission is to shed light on the origin and evolution of the early solar system by visiting the most massive objects in the Main Asteroid Belt, (4) Vesta and (1) Ceres. The Dawn spacecraft completed its mission at asteroid Vesta and will rendezvous with Ceres staring March 2015 (Russell and Raymond 2011). One of the Dawn mission's primary objectives is to determine the origin and evolution of Ceres by mapping its surface's mineralogical composition. Ceres is thought to be a water-rich object that has experienced significant aqueous processing (e.g., Lebofsky et al., 1981, Rivkin et al. 2003, Takir and Emery 2012, Kuppers et al. 2014).

Absorption features in spectra at ~3.0 µm are particularly indicative of aqueous alterations. These absorptions are likely due to hydroxyl- and/or water-bearing materials (Takir and Emery 2012), but could also be due to surficial OH implanted from solar wind (Sunshine et al. 2009) or exogenic sources like those seen on Vesta (Reddy et al. 2012a, Nathues et al. 2014). On the basis of the 3.0-µm absorption feature, Takir and Emery (2012) discovered that the surface composition of Ceres is not as unique as once thought. The authors found that asteroids (10) Hygiea and (324) Bamberga have both similar 3.0-µm absorption band centers (at 3.05 µm) and shapes as Ceres, suggesting that Ceres' surface mineralogy is not unique. Takir and Emery (2012) grouped Ceres, Hygiea, and Bamberga in one group on the basis of the 3.0-µm band center, and called it the Ceres-like group. In addition to the Ceres-like group, the authors discovered another group, Europa-like, which has a similar band shape as the Ceres-like group, but with a 3-µm band center at 3.14 µm.

Here, we analyzed Ceres' long wavelength cross-dispersed (LXD: 1.9-4.2 µm) spectra measured at phase angles ranging from $0.7^o$ to $22^o$. Our hypothesis is that phase angle variation of the observed asteroids affects the 3-µm band parameters, and hence affects the mineralogical interpretation of spectral features and diversity in asteroids, including Ceres. The 3-µm band parameters we used in this study include band center, band depth, and band area. Phase angle-induced spectral effects have been detected and quantified from laboratory measurements of powdered materials (e.g., Gradie et al. 1980; Gradie and Veverka 1982), ground-based observations of asteroids (e.g., Gehrels 1970; Millis et al. 1976; Lumme and Bowell 1981; Nathues, 2000, 2010; Reddy et al. 2012b; Sanchez et al. 2012), and spacecraft observations of asteroids (e.g., Clark et al. 2002; Bell et al. 2002; Kitazato et al. 2008). These effects can manifest themselves as phase reddening, which is characterized by increase spectral slope and variations in the strength of the absorption bands in visible-near infrared (VNIR) spectra. Therefore, our results will have implications on the interpretation of ground-based and Dawn spacecraft measured spectra of Ceres, as the 3-µm band center could be used for mineralogical characterization (e.g., Takir and Emery 2012) and identifying minerals in the regolith. Band depth has been used as as a proxy for the abundance or absence of OH or OH-bearing minerals on the surface (Sunshine et al. 2009).

## 2. Methodology

*1. Observational Techniques and Data Reduction*

Ceres spectra were collected between September 12, 2011 and December 18, 2012 at various phase angles ranging from $0.7^o$ to $22^o$, using the LXD (1.9-4.1 µm) mode of the SpeX spectrograph/imager at the NASA Infrared Telescope Facility (IRTF) (Rayner et al. 2013). Table 1 includes the Ceres observing parameters with LXD mode of SpeX.

We used standard Near-infrared (NIR) reduction techniques in order to reduce the LXD spectra. The data were reduced using the Interactive Data Language (IDL)-based spectral reduc-tion tool Spextool (v3.4) (Cushing et al. 2004) in combination with some custom IDL routines (Emery and Brown, 2003). The reduction techniques include subtracting the Ceres and standard star spectra at beam position A from the spectra at beam position B of the telescope in order to remove the background sky. Ceres and standard star spectra were extracted by summing the flux at each channel within an 8-pixel wide aperture. Ceres spectra were shifted to sub-pixel accuracy to align with the calibration star spectra and then were divided by appropriate calibration star spectra at the same airmass (±0.05) to remove telluric water vapor absorption features. More information about the observational techniques and data reduction can be found in Takir and Emery (2012).

2. *Thermal Excess Removal*

For Ceres LXD spectra, thermal excess is substantial at longer wavelengths (λ > 3.3 μm) and can affect the slope beyond 3.6 μm. The thermal excess is related to the surface temperature, which is a function of solar distance and other properties including albedo, surface roughness, and thermal inertia.

We removed the thermal excess in Ceres' spectra following the methodology described in Takir and Emery (2012) and references therein. The thermal excess is generally defined as:

$$\gamma_\lambda = \frac{R_\lambda + T_\lambda}{R_\lambda} - 1 , (1)$$

where $R_\lambda$ is the reflected flux at a wavelength $\lambda$, $T_\lambda$ is the thermal flux at a given wavelength, and the $R_\lambda + T_\lambda$ is the measured relative spectrum. To constrain the model thermal flux, we fitted the measured thermal excess with a model excess. Then, we subtracted the model thermal flux from the measured relative spectrum.

To calculate the thermal flux in the 3-μm region, we used the Harris' (1998) Near-Earth Asteroid Thermal Model (NEATM), which includes the beaming parameter $\eta$, to adjust the sur-face temperature to match the measured thermal flux (e.g., Harris and Lagerros 2002). We obtained the following model inputs: heliocentric and geocentric distances, geometric albedo (0.09: Li et al. 2010), and phase angle at time of Ceres observations from the Jet Propulsion Laboratory Horizon online ephemeris generator[1]. For the slope parameter, $G$, we used a default value of 0.15 from Bowell et al. (1989).

*3. Calculation of Band Center, Band Depth, and Band Area*

We used the following band parameters to analyze LXD spectra of Ceres: band center, band depth, and band area. Following a standard technique described by Cloutis et al. (1986), we isolated absorption features in the 3-μm region, and divided each absorption feature by a straight-line continuum in wavelength space. We determined the continuum by maxima at 2.95-3.00 μm and 3.20-3.25 μm. The band center was determined by applying a sixth-order polynomial fit to the central part of the feature at 3.2 μm. We calculated the band depth, $D_b$, using the following equation:

$$D_b = \frac{R_c - R_b}{R_c}, (2)$$

where $R_b$ is the reflectance at the band center and $R_c$ is the reflectance of the continuum at the band center (Clark and Roush 1984). We also calculated the band area by integrating the spectral curve below the straight-line continuum. To calculate band center uncertainties, we used an average of five measurements by varying the central part of the feature. For band depth and area uncertainties, we also used an average of five measurements, changing the positions of the continuum and band maxima at 2.95-3.00 μm and 3.20-3.25 μm. For each measurement, we slightly changed the positions of the central part of the feature and the positions of the continuum and band maxima in order to obtain representative uncertainties of the true error for band center, depth, and area. We calculated uncertainty for each parameter by the 2σ standard deviation, which represents variability from the average.

---

[1] http://ssd.jpl.nasa.gov/horizons.cgi

## 3. Results

Figure 1 shows the measured LXD spectra of Ceres on December 18, 2012, September 12, 2011, October, 11, 2011, and September, 21, 2012 at phase angles of $0.7^o$, $6^o$, $11^o$, and $22^o$, respectively. The dashed line at 3.05 μm represents the calculated band center for Ceres spectra by Rivkin et al. (2003) and Ceres-like group spectra (Hygiea and Bamberga) by Takir and Emery (2012). Table 2 shows the calculated 3-μm band parameters (band center, band depth, and band area) of Ceres observed at different phase angles ranging from $0.7^o$ to $22^o$. Table 2 also includes one Ceres observation that was made by Rivkin et al. (2003).

## 4. Discussion

Our calculated 3-μm band centers of Ceres are consistent with 3-μm band centers calculated for Ceres by Rivkin et al. (2003) (3.05 μm) and for the Ceres-like group calculated by Takir and Emery (2012) (3.05±0.01 μm). Ceres' spectra exhibit a distinct absorption feature centered at a wavelength of 3.05-μm that is superimposed on a broader absorption feature from 2.8 to 3.7 μm. Miliken and Rivkin (2009) attributed the 3.05 μm feature to hydroxide brucite and the broad absorption to magnesium carbonates and serpentines.

We did not note any significant change of the 3-μm band center with phase angles from $0.7^o$ to $22^o$. The band center slightly increases from 3.06 μm at lower phase angles ($0.7^o$ and $6^o$) to 3.07 μm at higher phase angles ($11^o$ and $22^o$). The band depth and area decrease by ~20% and ~25%, respectively, from lower phase angles ($0.7^o$ and $6^o$) to higher phase angles ($11^o$ and $22^o$). Pommerol & Schmitt (2008) also investigated the effect of phase angle variations on the strength of the $H_2O$ NIR absorption bands in hydrated minerals. They observed a decrease in band depth (~ 30%) and band area (40 %) with increasing phase angles in the range of ~80°-140° in the 3-µm absorption band of a smectite sample. However, for lower phase angles (~10°-80°), only small variations (~ 2%) of these parameters were seen. The difference between our results and those obtained by Pommerol & Schmitt (2008) could be attributed, at least in part, to the fact that Pommerol & Schmitt (2008) did not acquire data at phase angles lower than 10°, and therefore their measurements were not affected by the opposition effect, which is known to affect reflectance spectra at phase angles less than 5° (e.g., Gradie et al. 1980; Kitazato et al. 2008).

Because most Main Belt objects, including Vesta and Ceres, have a maximum excursion in solar phase angles of ~25°, we found small effects of the phase angles on the 3-µm band center for ground-based observations. The Dawn spacecraft is not anticipated to acquire any data at phase angles less than 7° (Reddy et al. 2015) because of trajectory and orbit constraints. Most Dawn data will be collected at phase angles between 20° and 80° (Reddy et al. 2015). Hence, our ground-based photometric analysis of the Ceres 3-µm band would be complementary to Dawn spacecraft data. While there are known calibration issues with the 3-µm region of VIR data, mitigation plans are already in place to remedy these issues. Additionally, the depth of the 3-µm feature on Ceres is deeper than the one on Vesta, making these calibration issues less of a problem.

## 5. Summary

We measured LXD spectra of Ceres at different phase angles ranging from 0.7° to 22° to analyze the effects of phase angles on the 3-µm absorption band. We found that: (1) the band center slightly increases from 3.06 µm at lower phase angles (0.7° and 6°) to 3.07 µm at higher phase angles (11° and 22°); (2) the band depth decreases by ~20% from lower phase angles to higher phase angles; and (3) band area decreases by ~25% from lower phase angles to higher phase angles. These ground-based photometric analyses of the Ceres 3-µm band will have impli-cations for interpreting Ceres' spacecraft spectra, which will be acquired by Dawn starting in spring 2015.

## AKNOWLEDGEMENTS

This research is supported by NASA Planetary Geology Geophysics grant NNX14AN35G (PI:Reddy) and NASA Planetary Mission Data Analysis grant NNX14AN16G (PI: Le Corre). The IRTF is operated by the University of Hawaii under Cooperative Agreement No. NCC 5-538 with the National Aeronautics and Space Administration, Office of Space Science, Planetary Astronomy Program. We would like to acknowledge NASA IRTF staff for their assistance with asteroid observations.

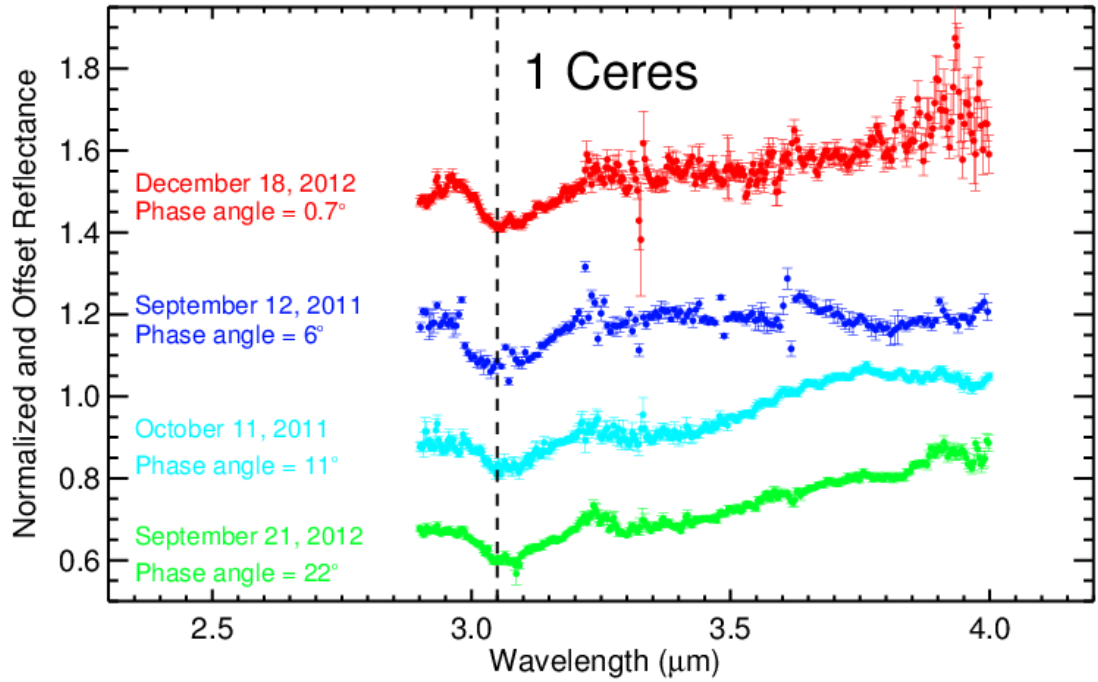

**Figure 1.** IRTF LXD Ceres spectra measured at different dates and phase angles ($0.7^{o}$ to $22^{o}$).

**Table 1.** Observing parameters for Ceres with LXD mode of SpeX.

| Date(UT) | Int (min) | Airmass | Standard star | Spectral type | B-V | V-K |
|---|---|---|---|---|---|---|
| Dec 18, 2012 | 50 | 1.0-1.4 | HD 252943 | G2V | 0.70 | 1.52 |
| Sep 12, 2011 | 40 | 1.3-2.1 | SAO147208 | G0V | 0.59 | 1.51 |
| Oct 09, 2011 | 40 | 1.3-1.4 | SAO147208 | G0V | 0.59 | 1.51 |
| Sep 21, 2012 | 20 | 1.0-1.1 | GSC 01863-02482 | G2V | 0.89 | 2.57 |

**Table 2.** The calculated 3-μm band parameters in spectra of Ceres observed at different phase angles ranging from 0.7° to 22°. [1]Spectra were measured and analyzed by Rivkin et al. (2003). Uncertainty for each band parameter in this study was determined by the 2σ standard deviation that represents variability from the average.

| | Phase angle (°) | 3-μm band center (μm) | 3-μm band depth (%) | 3-μm band area (μm$^{-1}$) |
|---|---|---|---|---|
| Ceres- Dec 18, 2012 | 0.7 | 3.06±0.01 | 12.81±0.68 | 0.02±0.01 |
| [1]Ceres- May, 2003 | 5 | 3.05 | -- | -- |
| Ceres- Sept 12, 2011 | 6 | 3.04±0.01 | 12.30±0.68 | 0.02±0.01 |
| Ceres- Oct 11, 2011 | 11 | 3.07±0.01 | 9.18±0.45 | 0.01±0.01 |
| Ceres- Sept 21, 2012 | 22 | 3.07±0.01 | 10.06±0.33 | 0.02±0.01 |